\documentstyle[12pt]{article}
\newcommand{\be}{\begin{equation}}
\newcommand{\ee}{\end{equation}}
\def\ba{\begin{eqnarray}}
\def\ea{\end{eqnarray}}

\def\({\left(}
\def\){\right)}

\begin{document}
\begin{center}

{\Large\bf 2D black holes and effective actions } \\

\vskip20mm

{\bf R. Balbinot\footnote{e-mail: balbinot@bo.infn.it} and A.
Fabbri\footnote{e-mail: fabbria@bo.infn.it}}

\end{center}

\vskip20mm

\noindent
Dipartimento di Fisica dell'Universit\`a di Bologna and INFN
sezione di Bologna, \\
Via Irnerio 46,  40126 Bologna, Italy

\vskip40mm

\begin{abstract}

We compare the canonical quantization and the effective action
method to derive expectation values of the stress energy tensor
for scalar fields conformally coupled to a 2D Schwarzschild black
hole spacetime. Particular attention is devoted to the thermal
equilibrium Hartle-Hawking state where the striking disagreement
of the results may be reconduced to the incomplete knowledge of
the effective action. We show how to reconcile the two procedures
and find physically meaningful analytical approximate expressions
for the stress tensor in the Hartle-Hawking state.

\end{abstract}
\newpage

\section{Introduction}

The study of quantum fields propagating in lower-dimensional spacetime is
usually regarded as an amusing playground which allows many interesting
features of ordinary 4D physics to be inferred by simple technical tools.
\\ \noindent
Within this perspective many efforts have been devoted to the
study of conformally coupled scalar fields propagating in a 2D
Schwarzschild spacetime \be \label{eu} ds^2=-(1-2M/r)dt^2 +
(1-2M/r)^{-1}dr^2 \>.  \ee These kind of studies are supposed to
give some hints on the quantum properties of the real 4D black hole at
least in the ``s-wave sector''. The basis objects of investigation
are the renormalized expectation values $\left< T_{\mu \nu}\right>$ of
the stress tensor operator for these quantum fields. 
These quantities can be calculated in
the canonical quantization scheme by mode sums and regularization
or, in a more elegant fashion, by functional variation of an
effective action. The latter method, requiring the knowledge of
the effective action for an arbitrary background metric, is far
more reaching for applications (see for example backreaction
calculations). It is obvious that the two procedures can not lead
to unequal results.
\\ \noindent
\par \noindent We first review the well
known Polyakov model \cite{Pol}, which describes massless and
minimally coupled 2d scalar fields. Its purpose is mainly
pedagogical, because in this simple context it is easy to show the
agreement of the two procedures, canonical quantization vs.
effective action. Particular attention however is needed to show
how thermal state expectation values can be retrieved by use of
effective action as emphasized by the authors of Ref.
\cite{fis}. We then turn our attention to a more sophisticated
theory, the so called dimensional reduction model, which has
received much interest in the literature \cite{va} (for a recent review see 
\cite{kuva}). Here things
become trickier since the results coming from canonical
quantization  \cite{bffnsz} do not seem to
match those coming from the effective action. It is widely
believed that such problems arise because of our incomplete
knowledge of the exact effective action (see for instance the
first of Refs. \cite{va} and also \cite{lmrz}, \cite{gzz}).
What is known is the so called anomaly induced 
effective action $S_{AI}$,
obtained by integrating the conformal anomaly. This procedure
allows the effective action to be known up to a Weyl invariant
functional. Based on this observation we propose two ways to
construct analytical expressions for the stress tensor in
agreement with the results inferred from canonical quantization.
The first is based on the anomalous transformation law of the
effective action under conformal transformations and is constructed 
using an ansatz \`a la Brown-Ottewill \cite{bop}. The second
method requires a minimal addition to $S_{AI}$ in the form of
a nonlocal Weyl invariant term. Finally, we state our
conclusions.

\setcounter{equation}{0}
\section{The ``Polyakov'' model}

Let us start our discussion by considering a conformal invariant
minimally coupled scalar field $\hat f$ whose action is

\be \label{pc} S_m = -\frac{1}{4\pi} \int d^2 x \sqrt{-g}(\nabla
\hat f)^2 \ee leading to the field equation

\be \label{deq} \Box \hat f=0 \>.\ee In the 2D Schwarzschild
spacetime of eq. (\ref{eu}) one can expand the field operator
$\hat f$ in terms of Eddington-Finkelstein modes $\{ e^{-iwu},
e^{-iwv} \}$ where \ba \label{efin} & &u = t-r^*\ , \ \ \ v=t+r^*
\\ & & r^*=r+2M\ln|r/2M-1|. \ea
This is the so called ``Boulware vacuum'' construction and leads
after renormalization to \ba \label{buva}  \left<B|
T^t_t |B \right> &=& \frac{1}{12\pi f}\left[ \frac{2M}{r^3} -
\frac{7M^2}{2r^4} \right]\ , \\  \left< B| T^r_r |B \right> &=&
-\frac{1}{12\pi f} \left[\frac{M^2}{2 r^4}\right]\ , \ea 
where $f=1-2M/r$. 
The $\left<B| T^r_t
|B\right>$ component vanishes identically. From the above
equations one recovers the usual trace anomaly \be \label{tran}
\left< B| T^a_a |B\right> =\frac{R}{24\pi}= \frac{M}{6\pi r^3}\ .\ee The Boulware
vacuum describes vacuum polarization outside a static spherically
symmetric body. Asymptotically this state coincides with the usual
Minkowski vacuum of flat space-time quantum field theory. However
it has a rather pathological behaviour at the event horizon.
Regularity of the stress tensor there in a regular frame requires
$(T^t_t - T^r_r)/f$ to be finite \cite{cf}. This condition is not fulfilled
by $\left< B| T^a_b|B\right>$. \par \noindent A quantum state
yielding a regular $T^a_b$ on the horizon can be obtained by using
a different set of modes as basis for the field operator. Choosing
the Kruskal modes $(e^{-iw U}, e^{-iwV})$ where \be \label{krmo}
U=-4Me^{-u/4M}\ , \ \ \ \ V=4Me^{v/4M}\ \ee one performs the so called
``Hartle-Hawking vacuum'' construction leading to
 \ba \label{hhva}  \left<H|
T^t_t |H \right> &=& \frac{1}{12\pi f}\left[ \frac{2M}{r^3} -
\frac{7M^2}{2r^4} \right] - \frac{1}{384\pi M^2 f}\ , \\  \left<
H| T^r_r |H \right> &=& -\frac{1}{12\pi f} \left[ \frac{M^2}{2 r^4}\right] +
\frac{1}{384\pi M^2 f} \ , \ea while as before \be  \left< H|
T^a_a |H\right> = \frac{M}{6\pi r^3}\ \ee which stresses the state
independence of the trace anomaly. One can check that
$\left<H| T^a_b |H\right>$ does
indeed satisfy the regularity condition on the horizon.
Asymptotically $\left<H| T^a_b |H\right>$ is not vanishing,
approaching the form \be \frac{\pi T_H^2}{6} \(
\begin{array}{cc}
 -1 & 0 \\
  0 & 1 \\
   \end{array} \) \ee
describing thermal radiation at the Hawking temperature
$T_H=1/(8\pi M)=1/2\pi\beta_H$. $|H > $ is indeed a thermal
state (the corresponding propagator being periodic in imaginary
time with period $2\pi \beta_H$) which describes a black hole in
thermal equilibrium with its radiation.\\ A better insight into
these results can be obtained by using the effective action
formalism. For the above theory the calculation of the effective
action is usually made by integrating the trace anomaly. The
result is the well known Polyakov action \cite{Pol} \be \label{po}
S_P=-\frac{1}{96\pi}\int d^2 x\sqrt{-g}R\frac{1}{\Box} R \>.\ee
However, as emphasized in Ref. \cite{fis}, writing the
effective action in the form (\ref{po}) one looses the information
about the state of the quantum field and in particular it is not
clear how (\ref{po}) can reproduce thermal radiation. \\ The
integration of the conformal anomaly does not give the absolute
value of the effective action, but rather the difference between
the effective actions for two conformally related
($g_{ab}=e^{2\sigma}\hat g_{ab}$) manifolds \be
\label{efconf}\Gamma (g)=\Gamma(\hat g)- \frac{1}{24\pi}\int d^2x
\sqrt{-\hat g} \left[ (\hat\nabla\sigma)^2 +\hat R\sigma \right]\ .\ee
The corresponding relation for the stress energy tensor is \be
\label{stct} T_{ab}(g)=T_{ab}(\hat g) - \frac{1}{48\pi} \left(
-4\hat\nabla_{\mu}\hat\nabla_{\nu}\sigma +4\hat\nabla_{\mu}\sigma
\hat\nabla_{\nu}\sigma + \hat g_{\mu\nu}(4\hat{\Box\sigma}
-2(\hat\nabla\sigma)^2)\right)\ .\ee By writing the 2D Schwarzschild metric
as \be ds^2=-(1-2M/r)(dt^2-dr^{*2})=-e^{2\sigma}d\hat s^2 \ee one
can take $\hat g$ as flat Minkowski space and setting $T_{ab}(\hat
g)=0$ one gets from eq. (\ref{stct}) \be T^t_t= \frac{1}{12\pi
f}\left[ \frac{2M}{r^3} - \frac{7M^2}{2r^4} \right]\ee which
coincides with eq. (\ref{buva}) and similarly for the $T^r_r$
component. So using as reference the usual Minkowski vacuum one
obtains as conformal image the Boulware state. \par\noindent To
obtain a thermal state at temperature $T$ ($=1/2\pi\beta$) one writes the
(Euclidean) Schwarzschild metric as \cite{fis} \be\label{inst}
ds^2=e^{2\sigma}d\hat s^2 \ee where now \be
e^{2\sigma}=\frac{\beta_H^2 f}{z^2}\ee and
\be
d\hat s^2=dz^2+(\alpha z)^2d\tilde\tau^2 \ ,\ee
\be \label{defo}
z=\beta_H e^{\frac{1}{\beta_H}\int \frac{dr}{f}}\ , \ee
where $\alpha= \beta/\beta_H$, $\tau=\beta \tilde \tau$ is
the euclidean time ($\tau=it$) $0\le \tilde\tau \le 2\pi$. \par
\noindent For $\beta=\beta_H$, $d\hat s^2$ is the metric of the
flat disk $D$ of radius $z_0$, whereas for $\beta\neq \beta_H$
$d\hat s^2$ is the metric of a flat cone $C_{\alpha}$ with deficit
angle $\delta=2\pi(1-\alpha)$. Let us first consider the regular
instanton ($\beta=\beta_H$). From eq. (\ref{stct}) we now obtain
\be \label{dh} T_{tt}= T_{tt}(D) + \frac{1}{12\pi }\left[-
\frac{2M}{r^3} + \frac{7M^2}{2r^4} \right] + \frac{1}{24\pi
\beta_H^2}\ . \ee Assuming the quantum field on the disk to be in
the state for which $T^a_b(D)=0$ we see that eq. (\ref{dh})
coincides with $\left<H|T^t_t|H\right>$ of eq. (\ref{hhva}).
Similarly for the $T^r_r$ component. So the conformal image of the
vacuum on the disk is the Hartle-Hawking state. \\ For the cone
($\beta\neq \beta_H$) one has \be
T_{tt}(C)=\frac{( 1-\alpha^2)}{24\pi \beta^2}  \ee and by eq.
(\ref{dh}) one has \be T_{tt}=\frac{1}{12\pi }\left[-
\frac{2M}{r^3} + \frac{7M^2}{2r^4} \right] + \frac{1}{24\pi 
\beta^2}\ \ee which asymptotically describes thermal radiation at
the temperature $T=1/2\pi \beta$. Note that only for
$\beta=\beta_H$ we have regularity of the stress tensor at the horizon. \par \noindent
One can calculate $T_{\mu\nu}$ directly in terms of the black hole
metric by using a local expression of $\Gamma(g)$ in terms of
quantities defined only with respect to the black hole metric
$g_{\mu\nu}$. Introducing an auxiliary field $\psi$ by \be
\label{cu} \Box\psi=R \ee one can write the effective action
$\Gamma[g]$ as \be \label{lapo}\Gamma(g)=-\frac{1}{48\pi}\int d^2 x\sqrt{-g} \left(
\frac{(\nabla\psi)^2}{2} +\psi R\right) + \Gamma_0 \ee where
$\Gamma_0$ is a conformally invariant functional. Neglecting
$\Gamma_0$ (we will come back to this point), 
the stress tensor is expressed in terms of $\psi$ as
\be \label{lama}
T_{\mu\nu} =\frac{1}{48\pi} \left( 2\nabla_{\mu}\nabla_{\nu}\psi
-\nabla_{\mu}\psi\nabla_{\nu}\psi
+g_{\mu\nu}(-2R+\frac{1}{2}(\nabla\psi)^2\right)\ .\ee For the
Schwarzschild metric eq. (\ref{cu}) can  be solved by \be
\label{hosf} \psi=-\ln (f) +br^* .\ee 
Note that the second term on the r.h.s. of eq. (\ref{hosf}) 
is a homogeneous solution of the auxiliary field equation (\ref{cu}). 
$b$ is an arbitrary (for the moment) constant. Inserting in eq. (\ref{lama}) 
one obtains
\be
T_{tt}=\frac{1}{12\pi }\left[- \frac{2M}{r^3} + \frac{7M^2}{2r^4}
\right] +\frac{b^2}{96\pi}\ .\ee The correct thermal behaviour at
infinity is obtained by setting $b=2/\beta$ reproducing all
previous results. Note that near the horizon ($r\to 2M$) \be\psi\to
\psi_c =-2(1-\frac{\beta_H}{\beta})\ln z \ee which is regular
for $\beta=\beta_H$, whereas for  $\beta \neq \beta_H$
$\psi_c$ coincides with the solution of the cone equation \be
\Box_c\psi_c=R_c, \ee where \be
R_c=2\frac{(1-\alpha)}{\alpha}\delta(z)\ .\ee  
So far we have neglected any contribution of $\Gamma_0$ to the stress tensor.
But being $\Gamma_0$ conformally invariant it contributes to the stress tensor
just with a conserved traceless tensor. These conditions plus time independence 
are so stringent that the resulting tensor must be proportional to 
\be \label{mbz}\frac{1}{f} \(
\begin{array}{cc} -1 & 0 \\ 0 & 1
\end{array} \) \ee
for a static black hole spacetime. \\ 
Consider now what has been the effect of adding the
homogeneous solution to the eq. (\ref{cu}). 
By adding a homogeneous
solution $\psi_0$ ($\Box\psi_0=0$), $T_{\mu\nu}(\psi)$ transforms 
as follows 
\be
T_{\mu\nu}(\psi)\to T_{\mu\nu}(\psi+\psi_0)=T_{\mu\nu}(\psi) +
[T_{\mu\nu}(\psi_0) -\frac{Rg_{\mu\nu}}{24\pi}]
+\frac{1}{48\pi}\left[ -\nabla_{(\mu}\psi\nabla_{\nu)}\psi_0 +
g_{\mu\nu}(\nabla\psi)(\nabla\psi_0)\right]. \ee Note that
$T^a_a(\psi +\psi_0)=T^a_a(\psi)$. The two additional tensors
are conserved and traceless giving  \be T^a_b = \frac{b^2}{96\pi f} \(
\begin{array}{cc} -1 & 0 \\ 0 & 1
\end{array} \) .\ee
Comparing this with eq. (\ref{mbz}), one realizes that one  
can simply neglect $\Gamma_0$ in eq. (\ref{lapo}) and add a homogeneous solution to
the auxiliary field eq. (as we did) obtaining the general expression for the 
stress tensor. 
The overall coefficient is fixed requiring the
thermal radiation behaviour at infinity (i.e. $b=2/\beta$).
So for the Polyakov theory one obtains the correct 
thermal contribution just by adding the homogeneous solution to
the auxiliary field eq. (\ref{cu}).

\setcounter{equation}{0}
\section{The dimensional reduction model: canonical quantization}

We come now to the second, more interesting but intriguing
example. Let us consider the following scalar dilaton action \be
\label{pd} S_m^{(2)}=-\frac{1}{4\pi}\int d^2 x \sqrt{-g}
e^{-2\phi}(\nabla \hat f)^2\>.\ee Unlike the previous case ( eq.
(\ref{pc}) ) the scalar field $\hat f$ is now coupled not only to 2D
gravity but also to a dilaton field. The action eq. (\ref{pd}) is
however still conformal invariant. This action can be obtained by
dimensional reduction of the following 4D action
\be
S_m^{(4)}=-\frac{1}{(4\pi)^2} \int d^4 x \sqrt{-g^{(4)}} (\nabla
\hat f)^2 \ee describing a minimally coupled 4D scalar field. The
reduction yielding eq. (\ref{pd}) is performed by assuming the 4D
metric to be spherically symmetric
\be
ds^2_{(4)}=g_{a b}(x^a)dx^a dx^b + e^{-2\phi (x^a)}d\Omega^2\>,
\ee where $a,b=1,2,  \ d\Omega^2$ is the line element of the
transverse unit sphere, $e^{-\phi}$ is the radius and $\hat f=
\hat f(x^a)$.
Because of this feature , the model of eq. (\ref{pd}) has acquired
considerable interest \cite{va}-\cite{gzz}. 
It is supposed to give a better hint on the
physics of a real 4D black hole when compared to the model based
on $S_m$ (\ref{pc}).\\  The presence of the dilaton in the action
(\ref{pd}) makes the field equation for $\hat f$ more complicated.
Instead of the simple D'Alembert equation (\ref{deq}) seen before,
we have now \be \label{emo} \nabla^a (e^{-2\phi}\nabla_a
\hat f)=0\>.\ee For the 2D Schwarzschild spacetime this eq. becomes
(writing $\hat f=e^{-iwt}R(r)/r$, where $r=e^{-\phi}$) 
\be \label{nvb} -\frac{d^2R}{dr^{*2}}
+\frac{2M}{r^3}(1-2M/r)R -w^2R=0 \ .\ee The canonical quantization
procedure starts by finding a complete set of solutions to the
above equation of motion. Plane waves are no longer solutions as
the effective potential acts as a reflecting barrier. Normal modes
of eq. (\ref{nvb}) are not known analytically in explicit form.
However from their asymptotic behaviours near the horizon \ba
 & & {\stackrel{\rightarrow}{R}}\sim e^{iwr^{\ast}}+
     {\stackrel{\rightarrow}{A}}(w)\, e^{-iwr^{\ast}} , \nonumber \\
 & & {\stackrel{\leftarrow}{R}}\sim \stackrel{\leftarrow}{B}
     (w)e^{-iwr^{\ast}} \label{ho} ,
\ea and at infinity, \ba
 & & {\stackrel{\rightarrow}{R}}\sim \stackrel{\rightarrow}{B}(w)
     e^{iwr^{\ast}} , \nonumber \\
 & & {\stackrel{\leftarrow}{R}}\sim
     e^{-iwr^{\ast}} + {\stackrel{\leftarrow}{A}}(w)\, e^{iwr^{\ast}}\ ,
     \label{in}
\ea where $A$ and $B$ are the reflection and transmission
coefficients (see Ref.  \cite{dew}), the following expressions for
$\left< T_{a b}\right>$ can be derived without recursion to any
regularization procedure \cite{bffnsz} \be \label{nhb} \left<
B|T_a^b|B\right>_{r\to 2M} \sim \frac{1}{384\pi M^2 f} \(
\begin{array}{cc} 1 & 0 \\ 0 & -1
\end{array} \) \>,\ee
\be \label{nhc} \left< H|T_a^b|H\right>_{r\to\infty} \sim
\frac{1}{384\pi M^2} \( \begin{array}{cc} -1 & 0 \\ 0 & 1
\end{array} \) \>, \ee where $a,b=t,r$ and $f=1-2M/r$. \\
\noindent Furthemore using a WKB approximation for the modes,
performing a large $w$ expansion and regularizing the stress tensor
by point splitting and then performing renormalization one can
obtain approximate analytic expressions for $\left< T_{a
b}\right>$ in the two states \cite{bffnsz}. For the Boulware (zero
temperature) state we have \be \label{lu} \left< B|T_{tt}|B\right>_{WKB}
 =-\frac{1}{2\pi }\left(-\frac{fM}{6r^3} +
\frac{M^2}{12 r^4} - \frac{f^2}{4r^2} 
\ln(\frac{m^2 f}{4\lambda^2})\right)\>,\ee
\be \label{ll} \left< B|T_{rr}|B\right>_{WKB}
 =\frac{1}{2\pi f^2}\left(-\frac{fM}{2r^3} -
\frac{M^2}{12 r^4} + \frac{f^2}{4r^2} 
\ln(\frac{m^2 f}{4\lambda^2})\right)\>,\ee
where $m^2$ is a
renormalization scale and $\lambda$ an infrared cutoff. 
These are fixed by requiring that for $M=0$ one recovers the Minkowski results,
namely $\left< T_{ab} \right>=0$. This yields $m^2=4\lambda^2$.
For the
thermal equilibrium state  we have 
\be \label{qq} \left< T_{tt}\right>_{WKB}
= \frac{\pi T^2}{6}-\frac{1}{2\pi }\left(-\frac{fM}{6r^3} +
\frac{M^2}{12 r^4} - \frac{f^2}{4r^2}( 2\gamma+ 
\ln[\frac{m^2 \beta^2 f}{4}])\right)\>,\ee
\be \label{qr} \left< T_{rr}\right>_{WKB}
= \frac{\pi T^2}{6f^2}+\frac{1}{2\pi f^2}\left(-\frac{fM}{2r^3} -
\frac{M^2}{12 r^4} + \frac{f^2}{4r^2} 
( 2\gamma +\ln[\frac{m^2\beta^2 f}{4}])\right) \>,\ee
where $\gamma$
is Euler number.  In both states the trace 
\be \label{traq} \left< T^a_a \right> = -\frac{M}{3\pi r^3}\ee 
is the same.
This is not an
approximation but an exact result, being just a realization of the
2D conformal anomaly which gives the stress tensor a state
independent anomalous trace proportional to the $a_1$ Seeley-De
Witt coefficient. For the theory described by the action
(\ref{pd}) it is \be \label{atr} \left< T^a_a\right> =
(24\pi)^{-1}[R-6(\nabla\phi)^2 +6\Box\phi ]\ee from which eq.
(\ref{traq}) follows. \\ The crucial problem with eqs.
(\ref{qq}) and (\ref{qr}) is the logarithmic term, 
which would imply the nonregularity on the horizon of
the Hartle-Hawking stress tensor ($T=T_H$) in a free falling
frame. Logarithmic divergences of this
kind are unfortunately a constant feature in approximations
schemes based on the WKB expansion in the case where the Ricci
tensor is nonvanishing  (see the analogous problem in 4D in Ref.
\cite{AnHiSa}).  This is an artifact of the WKB approximation
which breaks down near $r=2M$. As shown in Ref. \cite{bffnsz}
$\left< H|T_{\mu\nu}|H\right>$ is indeed regular on the horizon and
no $f^2\ln f$ terms are present.
Finally, on the horizon we have \be \label{klj} \left< H| T_{t}^{\
t}|H\right>_{r=2M}= \left< H| T_{r}^{\ r}|H\right>_{r=2M} =
-\frac{1}{48\pi M^2} \>. \ee Unfortunately the calculations of the modes are
rather involved and one has not been able to find till now an analytic expression
which smoothly matches the regular behavior on the horizon with the $r>>2M$
behavior of eqs. (\ref{qq}) amd (\ref{qr}).

\setcounter{equation}{0}
\section{The dimensional reduction model: effective action formalism}

The crucial question of this paper is how one can reproduce the
results of section 3 inferred by the canonical quantization
procedure using the effective action formalism and in particular
whether one can find an (at least approximate) analytical expression for the 
$T_{\mu\nu}$ in
the Hartle-Hawking state which has the required behaviour at
infinity and at the horizon. \\ By integrating the trace anomaly eq.
(\ref{atr}) one can write the effective action $\Gamma(g)$ for the
model in the following form \be \label{eac}
\Gamma(g)=-\frac{1}{2\pi}\int d^2 x \sqrt{-g}\left(
\frac{1}{48}R\frac{1}{\Box}R
-\frac{1}{4}(\nabla\phi)^2\frac{1}{\Box}R +\frac{1}{4}\phi
R\right) + \Gamma_0 = S_{AI}(g) + \Gamma_0(g) \ee where we have
indicated with $S_{AI}$ the anomaly induced contribution and
$\Gamma_0$ as usual is an arbitrary conformal invariant (not
necessarily local) functional. The difference of $S_{AI}$ with
respect to the Polyakov action are the two terms (one nonlocal and
the other local) containing the dilaton. \\ An interesting attempt to
find a workable expression of $\Gamma(g)$ including finite
temperature corrections has been made in Ref. \cite{BaFaNiSu}.
Using the high frequency approximation, a new approximation scheme
for quantum fields in (static) curved space, one has obtained the
following form of $\Gamma(g)$
\be
\Gamma_{HFA}(g)=\int d^2 x \sqrt{-g}\left[
-\frac{1}{24\pi\chi^2\beta^2} +\frac{(\nabla\eta)^2}{6\pi}
-\frac{1}{8\pi}\ln(\frac{m^2\beta^2\chi^2}{4 
e^{-2\gamma}})(\frac{R}{6}
-(\nabla\phi )^2+\Box\phi )\right] \ , \ee where
$\eta=-\frac{1}{4}\ln\chi^2$ and $\chi^2$ is the norm of the
Killing vector.\\  For zero temperature $\Gamma_{HFA}$ is just
$S_{AI}$ specialised to static 2D spacetimes.\\ Coming to the finite temperature case,
it is interesting to
note that neglecting the dilaton terms in $\Gamma_{HFA}(g)$, one has
an interesting expression for the finite temperature
Polyakov effective action, starting from which the results
of section 2 can be derived in an alternative and elegant way.\\ Coming back to
the actual problem, variation of $\Gamma_{HFA}$ with respect to
the metric gives
\ba
\left< T_{\mu\nu} \right>_{HFA} &=& \frac{1}{24\pi \beta^2\chi^2} \left[
g_{\mu\nu}-2\frac{\chi_{\mu}\chi_{\nu}}{\chi^2}\right]
-\frac{1}{6\pi}g_{\mu\nu}(\nabla\eta)^2 +
\frac{1}{3\pi}\nabla_{\mu}\eta\nabla_{\nu}\eta
-\frac{1}{6\pi}\frac{\chi_{\mu}\chi_{\nu}}{\chi^2}\Box\eta +\nonumber \\
& &\frac{1}{4\pi}\frac{\chi_{\mu}\chi_{\nu}}{\chi^2}\left[ \frac{R}{6} -
(\nabla\phi)^2+\Box\phi\right]
-\frac{1}{6\pi}[\nabla_{\mu}\nabla_{\nu}\eta -g_{\mu\nu}\Box\eta ]
-\frac{1}{2\pi}[\nabla_{\mu}\phi\nabla_{\nu}\eta + \nonumber \\ & &
\nabla_{\nu}\phi\nabla_{\mu}\eta - g_{\mu\nu}\nabla\phi\nabla\eta]
+\frac{1}{8\pi}\ln(\frac{m^2\beta^2\chi^2}{4e^{-2\gamma}})
[-g_{\mu\nu}(\nabla\phi)^2+2\nabla_{\mu}\phi\nabla_{\nu}\phi]\ . \ea 
For the Schwarzschild spacetime $\chi^2=(1-2M/r)$ and as shown in Ref. 
\cite{BaFaNiSu} 
these expressions 
coincide exactly with eqs. (\ref{qq}) and (\ref{qr}) obtained by
the canonical formalism under the large $w$ WKB approximation. So
the resulting $T_{\mu\nu}$ has the correct asymptotic behaviour, but
presents for $\beta=\beta_H$ the unphysical log divergence on the horizon. Therefore
$\Gamma_{HFA}(g)$ does not represent a valuable solution to our problem.
\par \noindent
Let us come back to the expression (\ref{eac}). For conformally
related metrics $g_{ab}=e^{2\sigma}\hat g_{ab}$ one has the
following relation which, we stress, is an exact identity
\be
\Gamma(g)=\Gamma(\hat g) -\frac{1}{24\pi}\int d^2 x\sqrt{-\hat g}
\left[(\hat\nabla\sigma)^2 +\hat R \sigma\right] -\frac{1}{4\pi}
\int d^2x \sqrt{-\hat g} \left[\sigma(\hat{\Box\phi} -
(\hat\nabla\phi)^2)\right]\ .\ee The corresponding relation for the
stress tensor is \ba T_{\mu\nu}(g) &=& T_{\mu\nu}(\hat g)-
\frac{1}{48\pi} \left[ -4\hat\nabla_{\mu}\hat\nabla_{\nu}\sigma
+4\hat\nabla_{\mu}\sigma \hat\nabla_{\nu}\sigma +\hat
g_{\mu\nu}(4\hat{\Box}\sigma
-2(\hat\nabla\sigma)^2)\right]+ \nonumber \\ & &\frac{1}{2\pi}\left[
\frac{\hat\nabla_{(\mu}\phi\hat\nabla_{\nu)}\sigma}{2}
-\frac{1}{2}\hat g_{\mu\nu}(\hat\nabla\phi)(\hat\nabla\sigma) +
\sigma
\hat\partial_{\mu}\phi\hat\partial_{\nu}\phi-\frac{1}{2}\hat
g_{\mu\nu}\sigma (\hat\nabla\phi)^2 \right] \ .\ea The idea is now to
proceed as in the Polyakov model (see section 2). Writing the
Schwarzschild metric as
\be
ds^2=-(1-2M/r)(dt^2-dr_*^2) \ee one gets \be T_{tt}(g)=T_{tt}(\hat g)+
\frac{1}{12\pi }\left[- \frac{2M}{r^3} + \frac{7M^2}{2r^4}\right]
+\frac{1}{2\pi }\left[- \frac{M f }{2r^3}+f^2 \frac{\ln f}{4r^2} \right]
\ee and similarly
\be
T_{rr}(g)= T_{rr}(\hat g) -\frac{1}{12\pi f^2}\left[ \frac{M^2}{2
r^4}\right] -\frac{1}{2\pi f}\left[ \frac{M}{2r^3}-f \frac{\ln f}{4r^2}
\right] \ee where $\hat g$ is Minkowski spacetime.\\ The striking
difference with the previous case is that here one is not allowed
to set $T_{ab}(\hat g)=0$ since it is true that $\hat g$ is flat
space, but our matter field $\hat f$ is now coupled not only to
the metric but to the dilaton as well ($\phi=-\ln(r)$) whose
expression in terms of $r^*$ is rather complicated. In particular,
from the anomaly one has \be T(\hat g)\equiv T^a_a(\hat
g)=\frac{1}{24\pi}(6\hat{\Box\phi}- 6(\hat\nabla\phi)^2)= -\frac{Mf}{2\pi r^3}\ee
which is indeed not vanishing. 
Writing
\be
T_{ab}(\hat g)=T_{ab}(g)-\frac{1}{2}\hat g_{ab} T(\hat g) +
\frac{1}{2}\hat g_{ab} T(\hat g) \ee and (\`a la Brown-Ottewill \cite{bop}) 
neglecting the traceless
contribution, we can write an approximate analytic expression for 
the zero temperature stress
tensor for the dimensional reduction theory as
\be \left< B|T_{tt}|B\right>_{BO}=-\frac{1}{2\pi}\left(\frac{M}{3r^3}-
\frac{7M^2}{12r^4}\right) +\frac{f^2 \ln(f)}{8\pi r^2}
\ ,\ee
\be
\left< B|T_{rr}|B\right>_{BO}=\frac{1}{2\pi f^2}\left( -\frac{fM}{r^3} 
-\frac{M^2}{12r^4}\right) 
+\frac{\ln(f)}{8\pi r^2}   \ . \ee As in eqs. (\ref{lu}), (\ref{ll}) we see that 
asymptotically  $\left< T_{\mu\nu}\right>\to 0$. Moreover, also the expression
(\ref{nhb}) is correctly reproduced. Despite these good properties, we see that 
 the polynomial parts in the
above eqs. are quite different from those we have obtained using the WKB
approximation scheme.
Performing the same calculation for the thermal state (using eqs.
(\ref{inst})-(\ref{defo}) ) we find
\be
T_{tt}(g)=T_{tt}(\hat g) +   \frac{f^2}{384\pi
M^2}\left[ 1+\frac{4M}{r} +\frac{60M^2}{r^2}-  48\frac{M^2}{r^2}\ln \frac{r}{2M}
\right] 
-\frac{Mf}{12\pi
r^3} \ , \ee
\be
T_{rr}(g)=T_{rr}(\hat g) + \frac{1}{384\pi
M^2}\left[ 1+\frac{4M}{r} +\frac{60M^2}{r^2}- 
 48\frac{M^2}{r^2}\ln \frac{r}{2M}\right]  +\frac{M}{12\pi
r^3 f} \ee where now $\hat g$ is the flat disc for
$\beta=\beta_H$ of the flat cone for $\beta\neq \beta_H$. Again
$T_{ab}(\hat g)\neq 0$. Neglecting as before the traceless part of 
$T_{ab}(\hat g)$ we can approximate the Hartle-Hawking stress tensor, \`a la
Brown-Ottewill,  as following 
\be \left< H|T_{tt}|H\right>_{BO} = \frac{f^2}{384\pi
M^2}\left[ 1+\frac{4M}{r} +\frac{60M^2}{r^2}
-48\frac{M^2}{r^2}\ln \frac{r}{2M}\right]  +\frac{f M}{6\pi
r^3}\ , \ee \be \left< H|T_{rr}|H\right>_{BO} =
 \frac{1}{384\pi M^2}\left[ 1+\frac{4M}{r}
+\frac{60M^2}{r^2} -48\frac{M^2}{r^2}\ln \frac{r}{2M}\right] 
-\frac{M}{6\pi r^3 f}\ . \ee
This stress tensor has extremely nice features.
 Comparison with $\left< H|T_{\mu \nu}|H\right>_{WKB}$ of
eqs. (\ref{qq})-(\ref{qr}) (with $T=T_H$) shows the correct asymptotic behavior not only in
the leading $T_H^2/6\pi$ term but also in the $1/r$ term which is quite
remarkable. Furthemore this $T^{\mu}_{\nu}$ has the property of being
regular on the horizon where the limiting value of eq. (\ref{klj}) are reached. 
This is our first proposal for $T_{\mu\nu}$ in
the Hartle-Hawking state. How good this approximation is cannot be
said because of our ignorance of the traceless part of
$T_{\mu\nu}(\hat g)$. Numerical computation  or a
better analytical approximation of $T_{\mu\nu}(\hat g)$ near the horizon are required.

\setcounter{equation}{0}
\section{The dimensional reduction model: local fields formulation}

As for the Polyakov case one can find a local expression for the
effective action depending on the background metric $g_{\mu\nu}$
and (now) two auxiliary fields $\psi$ and $\chi$ \cite{brm} \ba \label{lff}
\Gamma(g) &=& -\frac{1}{96\pi}\int d^2 x\sqrt{-g}\left[ 2R(\psi
-6\chi) +(\nabla\psi)^2
-12\nabla\psi\nabla\chi-12\psi(\nabla\phi)^2 +12R\phi\right]
+ \Gamma_0 \nonumber \\ &=& S_{AI}^{loc} + \Gamma_0\ea 
 where $\psi$ and $\chi$ satisfy \be
\label{taf} \Box\psi=R, \ \ \ \ \Box\chi=(\nabla\phi)^2\ ,\ee
$S_{AI}^{loc}$ is the local form of $S_{AI}$ and $\Gamma_0$ is an
arbitrary conformal invariant functional. If we neglect $\Gamma_0$
the stress tensor corresponding to $S_{AI}^{loc}$ is
\ba
\left< T_{\mu\nu}\right> &=& -\frac{1}{48\pi}\left[
2\nabla_{\mu}\nabla_{\nu}\psi -\nabla_{\mu}\psi\nabla_{\nu}\chi -g_{\mu\nu}(2R
-\frac{1}{2}(\nabla\psi)^2)\right] -\frac{1}{4\pi}[
-\frac{g_{\mu\nu}}{2}( (\nabla\phi)^2\psi + \nabla\psi\nabla\chi
 \nonumber \\ & -& 2(\nabla\phi)^2) +\psi \nabla_{\mu}\phi \nabla_{\nu}\phi +\frac{1}{2}
\nabla_{(\mu}\chi\nabla_{\nu)}\psi -\nabla_{\mu}\nabla_{\nu}\chi ]
+\frac{1}{4\pi}(g_{\mu\nu}\Box\phi - \nabla_{\mu}\nabla_{\nu}\phi)\ .\ea The
solution for the auxiliary field $\psi$ is (see eqs. (\ref{cu}) and
(\ref{hosf}) ) is \be \psi= -\ln f + br^*\ee where we set as before $b=2/ \beta$.
The other eq. (\ref{taf}) is solved by
\be
\chi=-\frac{1}{2}\ln(r/2M-1)-\frac{1}{2}\ln r/l \ . \ee The 
resulting stress tensor reads \ba \label{stco}
 T_{tt} &=& 
-\frac{1}{24\pi}(\frac{2M}{r^3}-\frac{3M^2}{r^4} - \frac{1}{
\beta^2}) -\frac{1}{2\pi}[ \frac{f^2}{4r^2}(\frac{2r}{\beta}
+\frac{4M}{\beta}\ln (r-2M)/l -\ln f )\nonumber \\ &+&\frac{1}{4M\beta
r^2}(-4Mr+4M^2)] +\frac{fM}{6\pi r^3} ] \ ,\ea
whereas 
\ba \label{rrcp}
 T_{rr} &=& 
-\frac{1}{24\pi f^2}(\frac{2M}{r^3}-\frac{3M^2}{r^4} - \frac{1}{
\beta^2}) -\frac{1}{2\pi}[ \frac{1}{4r^2}(\frac{2r}{\beta}
+\frac{4M}{\beta}\ln (r-2M)/l -\ln f )\nonumber \\ &+&\frac{1}{4M \beta
r^2 f^2}(-4Mr+4M^2)] -\frac{M}{6\pi r^3 f} ] \ .\ea
It is rather disappointing to see that in the regular instanton
case $\beta=\beta_H$, which would correspond to the
Hartle-Hawking case, the stress tensor 
\ba T_{tt}|_{\beta=\beta_H} &=&
 -\frac{1}{24\pi}(\frac{2M}{r^3}-\frac{3M^2}{r^4} -
\frac{1}{16M^2}) -\frac{1}{2\pi}[ \frac{f^2}{4r^2}(\frac{r}{2M} +\ln r/l
)\nonumber \\ &+& \frac{1}{32 M^2 r^2}(-4Mr+4M^2)]
+\frac{M f}{6\pi r^3} ]\ ,\ea 
\ba T_{rr}|_{\bar\beta=\beta_H} &=&
 -\frac{1}{24\pi f^2}(\frac{2M}{r^3}-\frac{3M^2}{r^4} -
\frac{1}{16M^2}) -\frac{1}{2\pi}[ \frac{1}{4r^2}(\frac{r}{2M} +\ln r/l
)\nonumber \\ &+& \frac{1}{32 M^2 r^2 f^2}(-4Mr+4M^2)]
-\frac{M }{6\pi r^3 f} ]\ ,\ea 
although having the correct asymptotic behaviour 
is not regular on the horizon.
One can
circumvent this negative result by including a homogeneous
solution to the second of eqs. (\ref{taf}). This, as shown in the
second of  Refs. \cite{brm}, allows to construct a $T_{\mu\nu}|_{\bar
\beta=\beta_H}$ which is regular on the horizon, unfortunately the
homogeneous solution modifies also the asymptotic behaviour giving
an unphysical negative Hawking radiation. The failure of the local
fields formalism to give meaningful results has to be found in the
completely arbitrary assumption we made to neglect $\Gamma_0$ in
(\ref{lff}). Unlike the previous case (Polyakov), now the inclusion of homogeneous
solutions to the auxiliary field equations does not reproduce completely 
the contribution coming from the unknown part of the effective action. The reason
is that now the corresponding $T_{\mu\nu}$ is not conserved (see eq. (\ref{noncons})
in Appendix A)
and the staticity and traceless
conditions are not sufficient to fix this contribution unambiguously.
Unfortunately for the moment no closed workable
expression for $\Gamma_0$ is known (see however the interesting
work of Ref. \cite{gzz}). In the following we will show how a simple form
of $\Gamma_0$ is sufficient to lead to a regular
$T_{\mu\nu}|_{\beta_H}$.

\setcounter{equation}{0}
\section{A ``phenomenological approach''}

Let us start by rewriting $S_{AI}^{loc}$ in a more symmetric form

\be
S_{AI}^{local}=S_1(\tilde \psi) + S_2(\tilde \chi) +S_{3}
\label{al} \ee where
\be
S_1(\tilde \psi)=\int d^2x\sqrt{-g}\left[ \frac{1}{2}\tilde
\psi\Box\tilde \psi
+\frac{1}{\sqrt{48\pi}}\left(R-6(\nabla\phi)^2\right)\tilde \psi
\right] \>,\label{un} \ee
\be
S_2(\chi)=\int d^2x\sqrt{-g}\left[-\frac{1}{2}\tilde
\chi\Box\tilde \chi+ \sqrt{ \frac{3}{4\pi} }(\nabla\phi)^2\tilde
\chi \right] \label{du} \ee and finally
\be
S_{3}=\int d^2x\sqrt{-g}\left[ -\frac{1}{8\pi}\phi R \right]
\>.\label{tr} \ee The auxiliary fields $\tilde \psi$ and $\tilde
\chi$ satisfy the equations of motion
\be
\Box\tilde
\psi=-\frac{1}{\sqrt{48\pi}}\left(R-6(\nabla\phi)^2\right) \>, \ee
\be
\Box\tilde\chi = \sqrt{ \frac{3}{4\pi} }(\nabla\phi)^2 \> . \ee
$\tilde\psi$ and $\tilde\chi$ are related to the fields $\psi$ and
$\chi$ of the previous section by
\be
\tilde \psi= -\frac{1}{\sqrt{48\pi}}(\psi - 6\chi), \ \ \ \tilde
\chi = \sqrt{\frac{3}{4\pi}}\chi\ .\ee 
The energy momentum tensor  can be
similarly splitted in three terms
\be
\left< T_{\mu\nu}\right> =  T_{\mu
\nu}^{1}(\tilde\psi) + T_{\mu\nu}^{2}(\tilde\chi) +T_{a b}^{3}
\label{ai} \ee where \ba T_{\mu\nu}^{1}(\tilde \psi)&=&
\partial_{\mu}\tilde \psi\partial_{\nu}\tilde \psi
+\frac{2}{\sqrt{48\pi}}\nabla_{\mu}\nabla_{\nu}\tilde \psi
+\frac{12}{\sqrt{48\pi}}\partial_{\mu}\phi\partial_{\nu}\phi \tilde\psi
\nonumber\\ &+&g_{\mu \nu}\left( -(\nabla\tilde\psi)^2
-\frac{2}{\sqrt{48\pi}}\Box\tilde\psi
-\frac{6}{\sqrt{48\pi}}(\nabla\phi)^2\tilde\psi \right)\>,  \ea
\ba T_{\mu\nu}^{2}(\tilde\chi)&=& -\partial_{\mu}\tilde\chi\partial_{\nu}\tilde\chi
-2\sqrt{\frac {3}{4\pi}}\partial_{\mu}\phi\partial_{\nu}\phi \tilde\chi
\nonumber\\ &+& g_{\mu \nu}\left( \frac{1}{2}(\nabla\tilde\chi)^2
+\sqrt{\frac {3}{4\pi}}(\nabla\phi)^2\tilde\chi \right) \>,  \ea
and
\be
T_{a b}^{3}= -\frac{1}{4\pi}(\nabla_{\mu}\nabla_{\nu}\phi- g_{\mu
\nu}\Box\phi) \>.  \ee One should note that $T_{\mu\nu}^{(2)}(\chi)$ is
traceless, being $S_2$ conformally invariant. On the other
hand we have that the trace $\left< T\right>\equiv \left<
T_{\mu}^{\mu}\right> $ is given by
\be
 T  \equiv \left< T_{\mu}^{\mu}\right> = T_{\mu}^{\mu 1}(\tilde \psi) 
 + T_{\mu}^{\mu 3}=
 -\frac{1}{\sqrt{12\pi}}\Box\tilde\psi + \frac{1}{4\pi}\Box\phi=
\frac{1}{24\pi}(R-6(\nabla\phi)^2 + 6\Box\phi)\>. \ee
Let us now give a closer look at the local form of $S_{AI}^{loc}$ eq.
(\ref{al}). 
We already remarked that the second nonlocal term is conformally invariant .
$\Gamma (g)$ is obtained by functional integration of the trace
anomaly. This procedure determines $\Gamma(g)$ up to conformal
invariant terms. Therefore terms like $S_2$, whose 
nonlocal expression is 
\be\label{lapop}
\frac{3}{8\pi} \int d^2x\sqrt{-g}(\nabla\phi)^2
\frac{1}{\Box}(\nabla\phi)^2 \>, \ee
 are completely
arbitrary, in particular its overall coefficient $\frac{3}{8\pi}$.
Given our present ignorance we shall try to mimic the effect of
this unknown part by adding to $S_{AI}^{loc}$ an additional
nonlocal term proportional to (\ref{lapop})
which being Weyl invariant
does not alter the trace anomaly. As a consequence the
nonlocal effective action we shall consider is the following \be \label{modp}
S_{AI}^{local} + \int
d^2x\sqrt{-g}\left(\frac{l_1^2}{2}-\frac{3}{8\pi}\right)(\nabla\phi)^2
\frac{1}{\Box}(\nabla\phi)^2 \>, \ee where $l_1$ is an arbitrary
parameter (see a similar procedure in 4D in \cite{bfs} ). This
should mimic the state dependence of the effective action, the
parameter $l_1$ taking different values according to the state of
the quantum field. In the local formulation $S_2$ is modified to
\be
S_2(\chi)=\int
d^2x\sqrt{-g}\left[-\frac{1}{2}\tilde\chi\Box\tilde\chi+
l_1(\nabla\phi)^2\tilde\chi \right] \label{dun} \ee and
accordingly
\be
T_{\mu \nu}^{2}(\chi)= -\partial_{\mu}\tilde\chi\partial_{\nu}\tilde\chi
-2l_1\partial_{\mu}\phi\partial_{\nu}\phi \tilde\chi  + g_{\mu \nu}\left(
\frac{1}{2}(\nabla\tilde\chi)^2 +l_1(\nabla\phi)^2\tilde\chi
\right) \>. \ee In the Schwarzschild spacetime the general sol. to
the auxiliary fields eqs. reads
\be
\tilde \psi=-\frac{1}{\sqrt{48\pi}}\left[ Ar + ( 2MA +2)
\ln(\frac{r}{2M}-1)
 +4\ln \frac{r}{l} \right]  \ee and
\be
\tilde \chi= \sqrt{\frac{3}{4\pi}}\left[ Br +(2MB
-\frac{1}{2})\ln(\frac{r}{2M}-1)
 -\frac{1}{2}\ln \frac{r}{l} \right] \>, \ee
where $A$ and $B$ are integration constants corresponding to the
homogeneous solution and $l$ is an arbitrary scale. 
 The corresponding expression for
$\left< T_{\mu \nu}\right>$ becomes (we shall work in $u,v$ 
Eddington-Finkelstein coordinates)
 \ba \left<T_{u u}\right> &=& \left<
T_{vv} \right> = \frac{1}{192\pi} \left(
fA+\frac{2AM+2}{r}+\frac{4f}{r}\right)^2 +\frac{1}{96\pi} \left(
\frac{2AM+2}{r^2} +\frac{4f^2}{r^2} \right) \nonumber \\
&-&\frac{f^2}{16\pi r^2}\left( Ar + (2AM+2)\ln(\frac{r}{2M}-1)
+4\ln \frac{r}{l} \right)\nonumber \\ &-&\frac{l_1^2}{4} \left(
+fB+\frac{2MB-1/2}{r}- \frac{f}{2r}\right)^2 \nonumber \\
&-&\frac{l_1^2 f^2}{2r^2} \left( Br
+(2MB-\frac{1}{2})\ln(\frac{r}{2M}-1) -\frac{1}{2}\ln \frac{r}{l}
\right) -\frac{f^2}{16\pi r^2} \> , \ea 
\be \left< T_{uv} \right> = \frac{1}{12\pi} (1-\frac{2M}{r})\frac{M}{r^3}\ .\ee
We now proceed to determine
these constants. \\ \noindent The Boulware vacuum is required to concide
with the Minkowski vacuum when $M=0$. Vanishing of the logarithmic
part
\be
\lim_{M\to 0} -\frac{1}{16\pi}\left[ 2\ln (\frac{r}{2M}-1) +4\ln
\frac{r}{l} \right] +\frac{l_1^2}{4} \left[ \ln ( \frac{r}{2M}-1)
+ \ln \frac{r}{l} \right] =0 \ee requires $l=2M$ and $l_1^2=3/4\pi
$ (i.e. the extra term in (\ref{modp}) vanishes)  . The resulting $\left< B|T_{ab}|B\right>$ is \be \left< B|
T_{uu}|B\right> = \left< B|T_{vv}|B\right> =\frac{1}{24\pi}\left[
-\frac{M}{r^3}+\frac{3M^2}{2r^4}\right] +\frac{1}{16\pi
}(1-\frac{2M}{r})^2\frac{1}{r^2} \ln(1-\frac{2M}{r}) \>,\ee
\be \left<B| T_{uv} |B\right> = \frac{1}{12\pi} (1-\frac{2M}{r})\frac{M}{r^3}\ .\ee
In $(t,r)$ coords. this corresponds exactly to $\left< B| T_{\mu\nu}|B\right>_{WKB}$ of
 eqs. (\ref{lu}), (\ref{ll}) 
once we fix $m^2=4\lambda^2$.
 \\ \noindent We come
now to the thermal Hartle-Hawking case. Regularity on the horizon
\footnote{Working in $u,v$
coords. we mention that regularity in a free falling frame in both the future and past
horizon is achieved by requiring $\left< T_{uu} \right>/f^2 <\infty$, 
$\left< T_{vv} \right>/f^2 <\infty$ and $\left< T_{uv} \right>/f <\infty$.}
of the log and polinomial parts require \ba\label{csa}2AM+2&=&
-8\pi l_1^2 (2MB-\frac{1}{2})\>,\nonumber
\\ (2AM+2)(AM+2)&=&24\pi l_1^2 (2MB-\frac{1}{2})^2\>. \ea The
asymptotic condition of a thermal bath at temperature $T_H=(8\pi
M)^{-1}$ yields \be \label{lulu} \frac{A^2}{192\pi} -\frac{l_1^2
B^2}{4} = (768\pi M^2)^{-1}=T_H^2/12\pi \>.\ee
These equations admit, as general solution, a unique value for
$l_1$ given by
  \be
l_1=\frac{1}{2\sqrt{\pi}}\> \ee and two possibilities for the
constants $A,B$, that is $A=-\frac{1}{M}$, $B=\frac{1}{4M}$ (i.e.
$\tilde \psi$ and $\tilde \chi$ regular at $r=2M$) and
$A=-\frac{1}{2M}$, $B=0$. Both cases give the same Hartle-Hawking
stress tensor \be \label{lum} \left< H|T_{uu}|H\right> = \left<
H|T_{vv}|H\right> = \frac{f^2}{768\pi M^2}\left[ 1+ \frac{4M}{r}
+\frac{36M^2}{r^2}(1-4\ln \frac{r}{l}) \right]\>,\ee
\be
\left< H|T_{uv}|H\right> =
\frac{1}{12\pi}(1-\frac{2M}{r})\frac{M}{r^3}\>.\ee $l$ remains as
an arbitrary scale. In $(t,r)$ coordinates these expressions become
(where we fix $l=2M$)
\be\label{cbm} 
\left< H|T_{tt}|H\right> = \frac{f^2}{384\pi M^2}\left[ 1+ \frac{4M}{r}
+\frac{36M^2}{r^2}(1-4\ln \frac{r}{2M})\right] 
+  \frac{fM}{6\pi r^3}\ , \ee
\be \label{cbn} \left< H|T_{rr}|H\right> = 
\frac{1}{384\pi M^2}\left[ 1+ \frac{4M}{r}
+\frac{36M^2}{r^2}(1-4\ln \frac{r}{2M}) \right]
-  \frac{M}{6\pi r^3 f}\ . \ee
\\ \noindent Let us compare eqs. (\ref{cbm}), (\ref{cbn}) 
with the analytical expression for $\left< H|T_{ab}|H\right>_{WKB}$
obtained by canonical quantization eqs. (\ref{qq}), (\ref{qr}) (with $T=T_H$).
Being by construction both expressions regular on the
horizon we have the correct limiting value of $\left< H|
T^t_t|H\right>$ and $\left< H|T^r_r|H \right>$ at $r=2M$ given by  (\ref{klj}).
Asymptotically the correct limiting behaviour has been imposed by our ``phenomenological''
construction. However one can see that also the $1/r$ term has the
correct coefficient as provided by the WKB approximation (see eqs.
(\ref{qq}), (\ref{qr})). This is definitely a nontrivial outcome of the model.
Finally, comparing with $\left< H|T_{\mu\nu}|H\right>_{BO}$ of section 4 
we see that the only difference are in the numerical coeffs. of the terms 
$\frac{1}{r^2}$ 
and $\frac{1}{r^2}\ln \frac{r}{2M} $ inside the square parenthesis. 

\setcounter{equation}{0}
\section{Conclusions}

The theoretical relevance of the model described by the action
(\ref{pd}) lies in its intimate connection to real 4D physics. It
is therefore crucial that the predictions made by the analysis of
this model be trustworthy. In this spirit any proposed effective
action $\Gamma(g)$ which for a Schwarzschild black hole does not
nicely reproduce at least the asymptotic behaviour described by
eq. (\ref{nhc}) and is regular on the horizon should be regarded
with suspicion. Using the conformal transformation law of the
effective action we were able to find, \`a la Brown-Ottewill
\cite{bop}, an approximation for the analytical expression of the 
 $\left< T_{\mu\nu}\right>$ in Boulware and
Hartle-Hawking states. Another expression can be obtained by adding
to $S_{AI}$ a conformal invariant functional (see eq. (\ref{modp}).
We mention that a term of this type
arises in the perturbative expansion of the effective action in
powers of $P=(\nabla\phi)^2-\Box\phi$ proposed in \cite{gzz}.
In particular we have shown that for the Boulware vacuum this
extra term vanishes and $S_{AI}$ reproduces exactly the WKB
form of $\left< B|T_{ab}|B\right>$. For the Hartle-Hawking state
the extra term is crucial to have the correct asymptotic behaviour
and regularity on the horizon at the same time. The expressions we
have proposed give the correct value of $\left< H|T_t^{\
t}|H\right>$ and $\left< H|T_r^{\ r}|H\right>$ on the horizon and
agree with the WKB form of $\left< H|T_{\mu\nu}|H\right>$ even to the
$O(1/r)$ term for $r\to \infty$. 
Finally, for the quantum states considered one can construct the
2D analogue of the 4D pressure term $\left< P\right>\equiv \left<
T^{\theta}_{\theta}\right>$ using the nonconservation equations
for the 2D stress tensor $\left< T_{ab}\right>$. The results are
given in Appendix A.

\setcounter{equation}{0}
\appendix

\setcounter{equation}{0}
\section{Calculation of the pressure terms }

For the theory of eq. (\ref{pd}) the stress tensor is not conserved
(\cite{bafa}, \cite{kuva}). Indeed $\left< T_{\mu\nu}\right>$ satisfies the
following nonconservation eqs.
\be\label{noncons}
\nabla_{\mu} \left< T^{\mu}_{\nu}\right> =-\frac{1}{\sqrt{-g}}\left<
\frac{\delta S}{\delta\phi}\nabla_{\nu}\phi \right> \>.\ee This is
nothing else but the 4D conservation eqs. $\nabla_{\mu}\left<
T^{\mu}_{\ \nu}\right>=0$,  where the 4D tangential pressure
in the local field formulation of section 6 
 is 
represented by \ba \left< P\right> &=& \frac{1}{8\pi r^2
\sqrt{-g}} \frac{\delta S}{\delta\phi}= \frac{1}{4\pi r^2}[ l_1
(\Box\phi \tilde\chi
+f\partial_r\phi\partial_r\tilde\chi)\nonumber
\\ &-&\frac{6}{\sqrt{48\pi}}(\Box\phi\tilde\psi +f
\partial_r\phi\partial_r\tilde\psi ) +\frac{R}{16\pi} ] \>. \ea
Its general expression in the Scharzschild spacetime is \ba \left<
P\right>&=& \frac{1}{4\pi r^2} [ l_1^2\frac{(1-\frac{4M}{r})}{r^2}
\left( B(r+2M\ln (\frac{r}{2M}-1)) -\frac{1}{2}\ln
(\frac{r}{2M}-1) -\frac{1}{2}\ln \frac{r}{l} \right) \nonumber
\\ &-&l_1^2\frac{f}{r}\left( B+\frac{2MB-1/2}{r-2M} -\frac{1}{2r}
\right)\nonumber
\\ &+&\frac{1-\frac{4M}{r}}{8\pi r^2}\left(   A(r+2M\ln(\frac{r}{2M}-1))
 +2\ln(\frac{r}{2M}-1) +4\ln \frac{r}{l} \right)\nonumber \\&-&\frac{f}{8\pi r}\left(A+
\frac{2AM+2}{r-2M} +\frac{4}{r}\right) +\frac{M}{4\pi r^3} ] \>.
\ea Substituting the values found for $A, B, l_1 $ in the states
considered we find $\left< P\right>$ in Boulware and
Hartle-Hawking states 
\be
\left< B| P|B\right> = \frac{1}{32\pi^2}\left[ \frac{4M}{r^5}
-\frac{(1-4M/r)}{r^4}\ln (1-\frac{2M}{r})\right] \>, \ee
\be
\left< H| P|H\right> = \frac{1}{32\pi^2} \left[ \frac{8M}{r^5}
-\frac{2}{r^4} + 3\frac{(1-\frac{4M}{r})}{r^4}\ln
\frac{r}{2M}\right] \>.\ee

 In particular, we find that $<B|P|B>$
exhibits a logarithmic divergence ($\sim \ln(1-2M/r)$) at the
horizon, whereas  $<H|P|H>$ is  regular there.
On the other hand substituting the $\left< T_{\mu\nu} \right>_{BO}$ approximations 
of section 4 one obtains 
\be
\left< B|P|B\right>_{BO}=\frac{1}{8\pi r}\left[ -(1-\frac{4M}{r})\frac{\ln(f)}{4\pi r^3}
+\frac{M}{4\pi r^4} +\frac{1}{2\pi f}\left( \frac{3M}{r^4}-\frac{7M^2}{r^5}\right)\right],
\ee
\ba
\left< H|P|H\right>_{BO}&=& \frac{1}{768 \pi^2 Mr^3} [ 1+\frac{4M}{r}+ 
\frac{108 M^2}{r^2}
-\frac{48M^2}{r^2} \ln \frac{r}{2M}\nonumber \\
 &-& f ( 1+ \frac{30 M}{r}
 -\frac{24M}{r}\ln\frac{r}{2M} + \frac{12M}{r} ) ]\ .\ea

\section*{Acknowledgements}
We wish to thank J. Navarro-Salas for useful discussions.

 \end{document}